\documentclass[12pt]{article}
\usepackage{amsmath,amssymb,hpkfinal}
\newcommand{\del}{\delta}
\title{Solving the Hamilton-Jacobi equation for gravitationally\\%
 interacting electromagnetic and scalar fields
\thanks{Supported in part by NSERC grant A8059.}
}
\author{
Bahman Darian\thanks{darian@phys.ualberta.ca}\\
Theoretical Physics Institute, University of Alberta\\
Edmonton, Canada T6G 2J1}
\begin{document}
\maketitle

\begin{abstract}

The spatial gradient expansion of the generating functional was recently
developed by Parry, Salopek, and Stewart to solve the Hamiltonian constraint 
in 
Einstein-Hamilton-Jacobi theory for gravitationally interacting dust and scalar
fields.  This expansion is used here
to derive an order-by-order
solution of the Hamiltonian constraint for gravitationally interacting
electromagnetic and scalar fields.  A conformal transformation and 
functional integral are used to derive the generating functional up to the 
terms
fourth order in spatial gradients.  The perturbations of a flat 
Friedmann-Robertson-Walker cosmology with a scalar field, up to second order 
in spatial gradients, are given.  The application of this formalism is 
demonstrated in the specific example of an exponential potential.  

\end{abstract}

\section{Introduction}
Hamilton-Jacobi (HJ) theory has many applications in the perturbative and 
non-perturbative
analysis of dynamical systems in classical mechanics.  Peres developed an 
Einstein-Hamilton-Jacobi (EHJ) formulation of general relativity in 
which a generating 
functional has to satisfy the momentum and Hamiltonian
constraints of general relativity \cite{peres}.
In the framework of quantum cosmology, it was known that the momentum 
constraints require any wave functional to be diffeomorphism invariant 
\cite{higgs}.  In a WKB approximation, such a requirement translates into
the diffeomorphism invariance of the generating functional.  
The Hamiltonian constraint is a non-linear functional partial
differential equation that governs the time evolution of the generating
functional.

Based on Peres' formalism,
Parry, Salopek, and Stewart used a series expansion of the generating 
functional in spatial gradients of the fields to derive an order-by-order
approximate 
solution of the Hamiltonian constraint for general relativity with matter
fields (see Ref. \cite{pss}, from now on referred to as
PSS).  Such a generating functional is diffeomorphism invariant in 
each order of the expansion.  Salopek and Bond used this formalism to show
how non-linear effects of the metric and scalar fields may be included in 
stochastic inflationary models.  The main advantage of 
this analysis is that the lapse function and shift vectors do not appear in 
the EHJ
equations.  Therefore, one obtains a coordinate-free approach to cosmological 
perturbations.  In the above models matter fields consist of self-interacting
scalar and dust fields.  

In this article the above formalism is extended to
minimally coupled gravitationally interacting scalar and electromagnetic fields.
Such minimally coupled electromagnetic fields give rise to conformally invariant
field equations.  Hence, the electromagnetic field energy density is 
proportional
to $1/a^4$ where $a$ is the scale factor.  Consequently, the electromagnetic 
field is diluted away during the De Sitter expansion phase of the inflationary 
cosmologies.  To break the conformal invariance, a direct coupling of gravity to 
electromagnetism \cite{turner} or corrections due to the 
quantum conformal anomaly
have been considered \cite{doglov}.
A coordinate-free approach to the perturbative analysis of cosmological 
models with electromagnetic fields could eventually lead to a better 
understanding of the primordial magnetic fields.  The generating functional 
up to 
the third
order in the spatial gradient expansion is given in section \ref{secI}.
Following PSS, 
section \ref{secII} is a demonstration of how a recursion relation and a 
functional integral 
in superspace can be used to derive the higher order terms in the 
spatial gradient 
expansion from the previous terms.  Section \ref{secIII} is an exhibition
of the gauge fixing and the solution of the field equations.
\section{ADM reduction and the EHJ equations}\label{secI}
The action for minimally coupled gravitationally interacting neutral scalar
and electromagnetic fields can be written as
\leqn{action}{
{\cal I}=\int\sqrt{\stackrel{(4)}{g}}[\half\stackrel
{(4)}{R}-
\half\phi_{,\mu}\phi^{,\mu}-V(\phi)-\quart F_{\mu\nu}F^{\mu\nu}]d^4x,\;
\mu=(0,1,2,3),
}
where $F_{\mu\nu}=A_{\nu,\mu}-A_{\mu,\nu}$ is the electromagnetic field 
strength and $V(\phi)$  is the scalar field potential.
ADM reduction of the above action is achieved by defining the 3-metric $
\gamma_{\mu\nu}:=g_{\mu\nu}+n_\mu n_\nu$ and the vector potential 
${\bf A}=
{\bf A}_{||}+{\bf A}_\perp$, such that 
$({\bf A}_{||})_\mu=\gamma^\nu_\mu A_\nu,\;({\bf A}_\perp)_\mu=-(n^\nu
 A_\nu)n_\mu$,
where $n_\mu$ is the unit vector field normal to space-like hypersurfaces of 
simultaneity parametrized by $t$.  In the basis 
$(\partial_t,\partial_i), i=1,2,3$,
such that $n^\mu(\partial_i)_\mu=0$ and $n^\mu=(\partial_t,-N^i\partial_i)/N$ 
(no sum), 
the
following relations hold:
$({\bf A}_{||})_i=\gamma_i^\nu A_\nu$, $A_0=N A_\perp+N^iA_i$, and
\leqn{metric}{
ds^2=-N^2dt^2+\gamma_{ij}(dx^i+N^i dt)(dx^j+N^j dt).
}
Then one follows with the procedure outlined in Ref. \cite{hanson} to derive 
the Lagrangian $L$ for gravitationally interacting electromagnetic and 
scalar fields.  With the momenta
$
\pi^{ij}=\Frac{\del L}{\del \dot{\gamma}_{ij}},\;\pi^\phi=
\Frac{\del L}{\del \dot{\phi}},\;E^i=
\Frac{\del L}{\del \dot{A}_i}, 
$
where $\dot{}:=d/dt$, after a Legendre transformation the Hamiltonian is
\leqn{hamiltonian}{
\mbox{Hamiltonian}=\int (N^\mu{\cal H}_\mu+A_0{\cal G})d^3x,
}
where
\leqn{Hconstraint}{ \begin{array}{rcl}
{\cal H}_0&=&\gamma^{-1/2}\pi^{ij}\pi^{kl}(2\gamma_{il}\gamma_{jk}-
\gamma_{ij}\gamma_{kl})+\gamma^{1/2}[V(\phi)-R/2+F^{il}F_{il}/4+\phi^{,i}
\phi_{,i}/2]\\
&+&
\gamma^{-1/2}[E^iE_i+(\pi^\phi)^2]/2=0,\;\mbox{Hamiltonian constraint},\\
{\cal H}_i&=&-2{\pi{_i^j}}_{|j}+F_{il}E^l+\pi^\phi\phi_{,i}=0,\;\mbox{
momentum constraint}, \\
{\cal G}&=&-{E^i}_{|i}=0,\;\mbox{Gauss law constraint}.
\end{array} }
${}_{|i}$ is the 3-covariant derivative (for covariant derivatives of 
tensor densities and sign conventions see \cite{mtw}).  Utilizing Hamilton's 
equations, the evolution equations for the fields are as follows:
\begin{eqnarray}
\dot{\phi} & = & N\gamma^{-1/2}\pi^\phi+N^i\phi_{,i}, \label{fieldphi} \\
\dot{\gamma}_{ij} & = & N2\gamma^{-1/2}\pi^{kl}(2\gamma_{il}\gamma_{jk}-
\gamma_{ij}\gamma_{kl})+2N_{(i|j)}, \label{fieldmetric} \\
\dot{A}_i & = & N\gamma^{-1/2}E_i+N^jF_{ji}+A_{0,i}. \label{fieldA}
\end{eqnarray}
The evolution equations for the momenta are considerably more complicated.
They are given by
\begin{eqnarray}
\dot{\pi}^\phi & = & -N\gamma^{1/2}\Frac{d V}{d \phi}-\half\gamma^{1/2}
\left(N_{,m}\phi^{,m}+N{\phi_{|l}}^{|l}\right)-\left(N^m\pi^\phi\right)_{,m}, 
\label{momenta1} \\
\dot{\pi}^{ij} & = & N\gamma^{-1/2}\left\{\gamma^{ij}\left[\pi^{mn}\pi_{mn}
-\half(\pi)^2\right]-4\pi^{im}\pi^j_m+2\pi^{ij}\pi \right.\nonumber \\
& + & \left. \quart\gamma^{ij}E^lE_l-\half
E^iE^j+\half\gamma^{ij}(\pi^\phi)^2\right\}-\half N\gamma^{1/2}\left\{
R^{ij}-\half\gamma^{ij}R \right. \nonumber \\
& - & \left. {F^i}_kF^{jk}+\quart \gamma^{ij}F_{lm}F^{lm}+\half\gamma^{ij}
\phi_{,l}\phi^{,l}-\phi^{,i}\phi^{,j}+\gamma^{ij} V(\phi)\right\} \nonumber \\
& + & \half\gamma^{1/2}
\left(N^{|ij}-{N_{|l}}^{|l}\gamma^{ij}\right)+\left(N^n\pi^{ij}\right)_{,n}
-2\pi^{n(i}{N^{j)}}_{,n}, \label{momenta2_2} \\
\dot{E}^{i} & = & \gamma^{1/2}\left(N_{,m}F^{mi}+N{F^{mi}}_{|m}\right)+
\left(N^mE^i-N^iE^m\right)_{,m}. \label{momenta3}
\end{eqnarray}

Instead of solving the evolution equations for the fields and momenta, 
one can try to solve the EHJ equations. The EHJ equations are 
derived by the substitutions 
\leqn{momenta2}{
\pi^{ij}=\Frac{\del S}{\del \gamma_{ij}},\;\pi^\phi=
\Frac{\del S}{\del \phi},\;E^i=\Frac{\del S}{\del A_i},
}
in ${\cal H}_\mu$ and $\cal G$.  $S=S[\gamma_{ij},\phi,A_i]$ is the 
generating functional (Hamilton's principal function)
\cite{goldstein}.  The Hamiltonian constraint is a 
hyperbolic functional partial differential equation for $S$.  
After solving the EHJ equations, (\ref{fieldphi}-\ref{fieldA}) and 
(\ref{momenta2})
yield the full set of the evolution equations. 
\section{The spatial gradient expansion and the order-by-order solution of the 
EHJ 
equations}\label{secII}
The momentum constraint implies that the generating functional
is diffeomorphism invariant \cite{peres}-\cite{higgs}.  One 
such-diffeomorphism invariant
quantity is $S=\int f[\phi,\gamma_{ij},A_i]\gamma^{1/2}d^3x$.  More generally,
a diffeomorphism invariant $S$ can be a multiple integral of some
multi-point functions.  The contribution of such  highly non-local terms could
be important, for example, if the spatial inhomogeneities are correlated.  
However, at least in the lowest orders, the contribution of such terms to the 
generating functional are expected to be insignificant in the generic case.  
Likewise, the Gauss law
constraint implies that $S$ is gauge-invariant, {\rm e.g.} $S=S[F_{ij}]$.  
Other gauge-invariant quantities like $\oint A_ldx^l$ could also be included 
in $S$. 
However, if the space-like surfaces of simultaneity are simply connected, 
one can write all such quantities in terms of $F_{ij}$ using Stokes
theorem.  Non-simply connected three-manifolds are not considered here.  The
Hamiltonian  constraint determines the time evolution of the fields.  
Following PSS, an order-by-order solution of the Hamiltonian constraint is
achieved by the expansion of the generating functional in spatial gradients:
\leqn{spatial}{
S={\DS\sum^\infty_{n=0}}\lambda^nS^{(n)}.
}
where $\lambda^{n}$ denotes the number of spatial gradients in $S^{(n)}$.  
It turns out that the scalar field is indispensable in this model and
dominates the dynamics of the space-time.  Moreover, 
in the limit $R\rightarrow 0$, in the absence of 
electromagnetism, the zeroth order 
solution is exact.  Therefore, merely based on dimensional grounds, 
this heuristically suggests the expansion parameter
should obey  
$\lambda\propto\Frac{\bar{R}}{V(\phi)}$.  In this relation, $\bar{R}$ 
is an appropriate 
combination of
the curvature invariants of dimension $L^{-2}$ 
such that in the flat space limit where the 
three-curvature 
is vanishing $\bar{R}=0$, and $V(\phi)$ represents the potential energy density
of the 
scalar field.  The convergence of the above series 
is an unsolved problem \cite{pss2}.

An order-by-order solution 
of the EHJ equation is achieved by substituting (\ref{spatial}) in the 
first equation in
(\ref{Hconstraint}) and the subsequent expansion of ${\cal H}_0$ in
spatial gradients:
\leqn{EHJ}{
{\cal H}_0=
{\DS\sum^\infty_{n=0}}\lambda^n{\cal H}^{(n)}
}
and requiring the EHJ equation to vanish at each order.  In the above equation
\leqn{H0}{
{\cal H}^{(0)}=\gamma^{-1/2}\Frac{\del S^{(0)}}{\del\gamma_{ij}}
\Frac{\del S^{(0)}}{\del\gamma_{kl}}(2\gamma_{il}\gamma_{jk}-\gamma_{ij}
\gamma_{kl})+\gamma^{1/2}V(\phi)+\gamma^{-1/2}
(\Frac{\del S^{(0)}}{\del \phi})^2/2=0
}
One can easily obtain the first few terms in (\ref{spatial}) by an ansatz.  The
zeroth order term
\leqn{lwa}{
S^{(0)}=-2\int\gamma^{1/2}H(\phi)d^3x,
}
called the long-wavelength approximation (LWA),  is the same as in 
PSS for some arbitrary function $H(\phi)$.  Inserting $S^{(0)}$ in 
(\ref{H0})
yields
\leqn{s0eqn}{
-3H^2+V(\phi)+2(\Frac{d H}{d \phi})^2=0. 
}
Electromagnetism has no dynamical degrees of freedom at this order.  
The LWA is very important in structure formation after inflation.  
Therefore, it is unlikely  that electromagnetic fields play a significant
role in structure formation in this model.    
For $n>0$
\leqn{HJeqn} {\begin{array}{rcl}
{\cal H}^{(n)} & = & \gamma^{-1/2}(2\gamma_{il}\gamma_{jk}-\gamma_{ij}
\gamma_{kl})\left(2\Frac{\del S^{(0)}}{\del 
\gamma_{ij}}\Frac{\del S^{(n)}}
{\del \gamma_{kl}}+{\DS\sum^{n-1}_{p=1}}\Frac{\del S^{(p)}}{\del \gamma_{ij}}
\Frac{\del S^{(n-p)}}{\del \gamma_{kl}}\right) \\
& + & \half\gamma^{-1/2}\gamma_{ij}{\DS\sum^{n-1}_{p=1}}\Frac{\del S^{(p)}}
{\del A_i}
\Frac{\del S^{(n-p)}}{\del A_j}+\gamma^{-1/2}\Frac{\del S^{(0)}}
{\del \phi}\Frac{\del S^{(n)}}{\del \phi}
+\half\gamma^{-1/2}{\DS\sum^{n-1}_{p=1}}\Frac{\del S^{(p)}}
{\del \phi}\Frac{\del S^{(n-p)}}{\del \phi}+\nu^{(n)}, \\
\nu^{(2)} & = & \gamma^{1/2}(-R+\phi^{,i}\phi_{,i}+F_{il}F^{il}/2)/2, \\
\nu^{(n)} & = & 0,\mbox{ for }n\not=2.
\end{array} }
At each order, ${\cal H}^{(n)}=0$ is a linear hyperbolic functional 
differential equation in the unknown functional $S^{(n)}$.

$S^{(2)}$ is the next non-vanishing
term given by
\leqn{S2}{
S^{(2)}=\int\gamma^{1/2}(J(\phi)R+K(\phi)\phi_{,i}\phi^{,i}+L(\phi)
F_{ij}F^{ij})d^3x. 
}
The above terms are not the only terms quadratic in spatial gradients in 
$S^{(2)}$.  However,
the remaining terms are either equal to the above terms modulo surface 
integrals or vanish identically.  For example, $\int\gamma^{1/2}F^{ij}
\epsilon_{ijk}\phi^{,k}d^3x=-\int\gamma^{1/2}(F^{ij|k}\epsilon_{ijk}\phi)
d^3x=0$ due to Maxwell equations $F_{[ij|k]}=0$.  It is easy to verify 
that $S^{(2)}$ satisfies the momentum and Gauss law
constraints.  One notices that 
$\Frac{\del S^{(2)}}{\del A_i}$ is already quadratic in spatial derivatives and
does not appear in the second order EHJ equation ${\cal H}^{(2)}=0$.  
After inserting $S^{(2)}$ in the second order EHJ equation and grouping 
together the coefficients of $R$, ${\phi_{|l}}^{|l}$, $\phi^{,l}\phi_{,l}$, and 
$F_{ik}F^{ik}$ one respectively has
\leqnarr{s2eqn}{
s_1:=HJ-2\Frac{d H}{d \phi}\Frac{d J}{d \phi}-
\Frac{1}{2}=0,& s_3:=HK-4J\Frac{d^2J}{d\phi^2}+2\Frac{dH}{d\phi}
\Frac{dK}{d\phi}
+\Frac{1}{2}=0,  \nonumber\\
s_2:=-H\Frac{d J}{d \phi}+K\Frac{d H}{
d \phi}=0, & s_4:=HL+2\Frac{d H}{d \phi}\Frac{d L}{d \phi}-\Frac{1}{4}=0. 
\label{s2eqn}
}
At first sight, (\ref{s2eqn}) seems to be an over-determined system for three
unknown functions, $J$, $K$, and $L$.  However, by solving $s_2=0$ for $K$, 
one can 
show that $s_3$ is not independent and obeys the relation 
$s_3=2\Frac{ds_2}{d\phi}-s_1+
\Frac{ds_1}{d\phi}H\left(\Frac{dH}{d\phi}\right)^{-1}$.  
In the spatial gradient expansion of the generating functional $S$ for the
scalar fields in the absence of electromagnetism, there is no
contribution from odd order terms.  Electromagnetism makes non-trivial 
contributions to odd orders.  The only non-vanishing term in $S^{(3)}$ is
\leqn{s3}{
S^{(3)}=\int\gamma^{1/2}M(\phi){F^{ij}}_{|j}\epsilon_{ikl}F^{kl}d^3x.
}
All other third order terms like ${F^{ij}}_{|j}\phi_{,i},\;F^{mn}F^{lp}
\epsilon_{mnl}\phi_{,p},\;{F^{ij}}_{|ij}$ either vanish or are total 
divergences.  To solve the third-order EHJ equation ${\cal H}^{(3)}=0$
one has to use the relation $\Gamma^i_{ji}=\gamma^{-1/2}(\gamma^{1/2})_{,i}$.
The solution yields
\leqn{s3eqn}{
MH+\Frac{d M}{d \phi}\Frac{d H}{d \phi}=0.
}
$S^{(3)}$ non-trivially satisfies the momentum and Gauss law constraints.  To
show this, one frequently uses the identity $V_s\epsilon_{ijk}=V_i
\epsilon_{sjk}+V_j\epsilon_{isk}+
V_k\epsilon_{ijs}$ for any vector field $V_i$, achieved from $\Frac{\del}
{\del \gamma_{rs}}\left(\epsilon^{ijk}\right)=\Frac{\del}{\del \gamma_{rs}}
\left(\gamma^{il}\gamma^{jm}
\gamma^{kn}\epsilon_{lmn}\right)$ and multiplication of both sides by $V_r$.  
Equations (\ref{s0eqn}), (\ref{s2eqn}), and
(\ref{s3eqn}) form a set of differential equations, easily solvable 
for
most relevant potentials.  Nevertheless, the full set of differential equations
 becomes 
increasingly complicated at higher orders.  
As in PSS one can use the expression for $S^{(0)}$ and the 
conformal transformation 
\leqn{conformal1}{f_{ij}=\gamma_{ij}\Omega^{-2}(u),\;  
u:=\int(-2\Frac{d H}{d \phi})^{-1}d\phi,\;
\Frac{d \Omega}{d u}=H\Omega,
}
to solve the EHJ equations (\ref{HJeqn}).  The EHJ equations transforms into
\leqn{conformal2}{
\Frac{\del S^{(n)}}{\del u(x)}\left|_{f_{ij},A_i}\right.=-\tilde{\cal R}^{(n)}
}
where
\leqn{conformal3}{\begin{array}{rcl}
\tilde{\cal R}^{(n)}&=&f^{-1/2}\Omega^{-3}(2f_{il}f_{jk}-f_{ij}f_{kl})
{\DS\sum^{n-1}_{p=1}}\Frac{\del S^{(p)}}{\del f_{ij}}\Frac{\del S^{(n-p)}}
{\del f_{kl}}\\
&+&
\Frac{f^{-1/2}}{8\Omega^{3}}\left(\Frac{dH}{d\phi}\right)^{-2}
{\DS\sum^{n-1}_{p=1}}(\Frac{\del S^{(p)}}{\del u}
-2\Frac{\del S^{(p)}}{\del f_{lm}}f_{lm}H)(\Frac{\del S^{(n-p)}}{\del u}\\
&-&
2\Frac{\del S^{(n-p)}}{\del f_{pq}}f_{pq}H)+\Frac{f^{-1/2}}{2\Omega}f_{ij}
{\DS\sum^{n-1}_{p=1}}\Frac{\del S^{(p)}}{\del A_i}\Frac{\del S^{(n-1)}}{\del A_j}+
\tilde{\nu}^{(n)}\\
\tilde{\nu}^{(2)} & = & f^{1/2}\{\Frac{\Omega}{2}\tilde{R}+\Frac{d}
{d u}
(H\Omega)u_{,i}u_{,j}f^{ij}-\Frac{1}{4\Omega}F_{ik}F_{lm}f^{il}f^{km}\},
\;\tilde{\nu}^{(n)}=0, \mbox{for}\;n\not=2.
\end{array} }
$\tilde{R}$ is the conformal curvature and all indices are raised and lowered 
with the conformal metric $f_{ij}$.

The proof of the integrability of (\ref{conformal2}) for gravitationally
interacting electromagnetic and scalar fields proceeds similar to what was
outlined in \cite{salopek1}.  By using the expression for $\tilde{\nu}^{(2)}$ from
(\ref{conformal3}), a functional integration of (\ref{conformal2}) 
gives rise to the following expression for $S^{(2)}$:
\leqn{s2conformal}{
S^{(2)}=\int f^{1/2}\left(j(u)\tilde{R}+k(u)u_{,i}u^{,i}+l(u)F_{ik}F^{ik}\right)d^3x,
}
where
\leqn{s2conformal2}{
j(u):=\int^{u}_0\Frac{\Omega(u')}{2}du'+D,\;l(u):=-\int^u_0\Frac{du'}
{4\Omega(u')}+D',\;k(u):=H\Omega.
}
The complementary functionals $D$, $D'$ are constants of integration.
In the next order $\Frac{\del S^{(3)}}{\del u}=0$, therefore 
$
S^{(3)}=\int f^{1/2}{F^{ij}}_{;j}\varepsilon_{ikl}F^{kl}d^3x
$
is the most general form of $S^{(3)}$ in which $;$ 
 and $\varepsilon_{ijk}$ are the covariant derivative and Levi-Civita tensor 
associated with 
$f_{ij}$ respectively.  Conformal transformation of this expression gives 
rise to
\leqn{s3conformal}{
S^{(3)}=\int \gamma^{1/2}\Omega^2 {F^{ij}}_{|j}\epsilon_{ikl}F^{kl}d^3x.
}
Functional integration of (\ref{conformal2}) in the next order gives
rise to
\leqn{s4conformal}{
\begin{array}{rcl}
{\cal S}^{(4)} & = & \int d^3x f^{1/2}\{-\ell(u)\tilde{R}^{ij}\tilde{R}_{ij}
-(3\ell(u)/8+m(u))\tilde{R}^2
-n(u)(\tilde{R}^{ij}-f^{ij}\tilde{R}/2)u_{,i}u_{,j}
 \\
& + &r(u)u_{;l}^{;l}u^{,j}u_{,j}+s(u)(F_{lm}F^{lm})^2+t(u)u_{,p}u^{,p}F_{lm}F^{lm}
+v(u)u_{,m}u^{,n}F^{mi}F_{ni}\\
& + &w(u)u_{,m}F^{mi}{F^n}_{i;n}+x(u) {F^{mi}}_{;mn}{F^n}_{i}+
y(u)F_{km}{F_l}^mF^{kn}{F^l}_n
+z(u)\tilde{R}
F_{ln}F^{ln}\\
& + & a(u)\tilde{R}_{kl}F^{km}{F^l}_{m}\}.
\end{array} }
where $\ell(u),m(u),n(u),r(u)$ are given in PSS.  In the above expression
 $s(u),\cdots,a(u)$ are defined as
\leqn{s4conformal2}{ \begin{array}{ll}
s'(u)=\Omega^{-3}[-11 l^2/4+(\Frac{d H}{d \phi})^{-2}(
\Omega^{-1}/4+Hl)^{2}/8], &
a'(u)=8jl\Omega^{-3},\\
z'(u)=\Omega^{-3}\left(-\Frac{5}{2}jl+\quart(\Frac{d H}{d \phi})^{-2}
(1/8+Hl\Omega/2-Hj\Omega^{-1}/4-H^2jl)\right), &x'(u)=-8l^2\Omega^{-1},
 \\
v'(u)=\Omega^{-2}(-4lH+3\Omega^{-1}/2),&y'(u)=8l^2\Omega^{-3}, 
 \\
w'(u)=4l\Omega^{-1}(2lH+\Omega^{-1}), & \\
t'(u)=\Omega^{-2}(Hl-\Frac{\Omega^{-3}}{3}),\\
\end{array} }
in which $':=d/du$.
\section{The evolution equations of the fields}\label{secIII}
Once the EHJ equations are solved, the evolution equations for the fields are
obtained from (\ref{fieldphi}-\ref{fieldA}) and (\ref{momenta2}).  A judicious
choice of gauge greatly simplifies the field equations.  In the almost
synchronous gauge ($N^i=0$) if $u$ is the time
parameter, from (\ref{fieldphi}) it follows that in the LWA ({\rm i.e} 
$S=S^{(0)}$) the lapse obeys $N^{(1)}=1$.  The superscript 
$(n)$ means that the right hand side of the equation contains terms up to 
$(n-1)$th order in spatial gradients.  

The choice of $u$ as the
time parameter is valid as long as the geometry is sufficiently close to that
of the homogeneous models (for a relevant discussion see Ref. 
\cite{ellis_bruni}).  
Then it is useful
to replace (\ref{fieldmetric}) with the equivalent evolution equation 
\leqn{fieldconformal}{
\dot{f}_{ij}=2N\Omega^{-3}f^{-1/2}\Frac{\del S}{\del f_{kl}}(2f_{jk}f_{il}-
f_{ij}f_{kl})-2Hf_{ij}
}
for the conformal metric $f_{ij}$ which is related to
$\gamma_{ij}$ via
\leqn{3metric}{
\gamma_{ij}=\mbox{\rm exp}\left\{-\int(\Frac{d H}{d \phi})^{-1}
Hd\phi\right\}f_{ij}. 
}
In the LWA $\dot{f}^{(1)}_{ij}=0$.  $f^{(1)}_{ij}$ is the seed metric that contains
no dynamical degrees of freedom.  The first non-trivial evolution equation for
$\dot{A}_i$ in the temporal gauge $A_0=0$, is obtained from the conformal 
transformation of (\ref{fieldA}):
\leqn{Aevolution1}{
\dot{A}^{(3)}_i=Nf^{-1/2}\Omega^{-1}f_{li}\Frac{\del S^{(2)}}{\del A_l}=
-4\Omega^{-1}{F^m}_{i;m}.
}
Since $d/du$ and $;$ commute, the evolution equation for $F_{ij}$ is easily
derived from the above equations to be
\leqn{Fevolution1}{
\dot{F}^{(3)}_{ij}=8\Omega^{-1}l(u){F^m}_{[j;i]m},
}
with the solution
\leqn{Fevolution2}{
F^{(3)}_{ij}=\mbox{\rm exp}\left\{-\left(8\int du\;\Omega^{-1}l(u)\right)
\del^k_{[i}\del^l_{j]}
\nabla_{k}\nabla^{m}\right\}
{\cal F}_{ml}. }
In the above equation and what follows, ${\cal F}_{ij}$ is an arbitrary
antisymmetric tensor field, $\nabla$ refers to the covariant derivative with
respect to 
the seed metric and the indices are raised 
with the seed metric.  The exponential of the matrix differential
operator is defined as:
\begin{eqnarray}
\mbox{{\rm exp}}\left\{\cdots\right\}&:=&\left\{\del^m_i\del^l_j-\left(8\int
\Omega^{-1}l(u)\;du\right)
\del^k_{[i}\del^{l}_{j]}\nabla_k\nabla^m \right.\\
&+&\left.\Frac{1}{2!}\left(8\int\Omega^{-1}l(u)\;du\right)^2
\del^p_{[i}\del^q_{j]}\nabla_p\nabla^r\del^k_{[q}\del^l_{r]}\nabla_k\nabla^m+
\cdots\right\}{\cal F}_{ml}.
\end{eqnarray} 
Once the evolution equations for the fields are solved, the evolution 
equations for 
the momenta are easily derived from (\ref{momenta2}).  In particular, the 
electric field obeys the equation
\leqn{electric}{
E^{(3)i}=\Frac{\del S^{(2)}}{\del A_i}=-4f^{1/2}l(u){F^{mi}}_{;m}
}
In higher orders (\ref{fieldphi}) shows that
$N^{(n)}\not=1$.  For example,
\leqn{n3}{
N^{(3)}=1-\Omega^{-3}\left[\tilde{R}^{(1)}\left(\Frac{d j}{d u}-jH\right)
+{\cal F}_{kl}{\cal F}_{mn}f^{(1)km}f^{(1)ln}\left(lH+\Frac{d l}{d u}\right)\right]
\left(-2\Frac{d H}{d \phi}\right)^{-2}
}
where $\tilde{R}^{(1)}$ is the three-curvature associated with the seed metric
$f^{(1)}_{ij}$.
Obviously, the higher order evolution equations are more complicated.  
The third order evolution of the conformal metric is derived from 
(\ref{fieldconformal}) and (\ref{n3}) to obey 
\leqnarr{3metricconf}{
\dot{f}^{(3)}_{ij} & = & \Omega^{-3}\left\{ \tilde{R}^{(1)}f^{(1)}_{ij}
\left[\Frac{H}{2}(\Frac{d H}{
d \phi})^{-2}(jH-\Frac{d j}{d u})+j\right]
-4j\tilde{R}_{ij}^{(1)}\right.\nonumber \\
& + & \left.{\cal F}_{kl}{\cal F}_{mn}f^{(1)km}f^{(1)ln}f^{(1)}_{ij}
\left[-\Frac{H}{2} (\Frac{
d H}{d \phi})^{-2}(lH+\Frac{d l}{d u})+
3l\right]-8l{\cal F}_{in}{\cal F}_{jm}f^{(1)mn}\right\}.\label{3metricconf}
}

As a demonstration of an application of the formalism developed so far,
one could compute $\dot{f}^{(3)}_{ij}$ for an arbitrary magnetic field
${\cal F}_{kl}$ and seed metric $\dot{f}^{(1)}_{ij}$ with a scalar potential
$V=V_0{\rm exp}\left\{-\sqrt{\Frac{2}{p}}\phi\right\}$.  The general solution
of (\ref{s0eqn}) for $p\not=1/3$ is given in \cite{salopek-bond}.  The general
parametric solution of (\ref{s0eqn}) with $H$ and $\phi$ as functions of 
an independent variable $v$ is, 
\leqnarr{pthree}{
H=\left[\Frac{V_0}{3}\right]^{\half} \exp\{-\phi\sqrt{\Frac{3}{2}}\}\cosh{v},
\\
\phi=\phi_m+\sqrt{\Frac{3}{2}}
\left(\pm \Frac{v}{2}+\Frac{e^{\pm 2v}}{4}\right),
}
where $\phi_m$ is the integration constant.
A special solution of (\ref{s0eqn}) for $p\not=1/3$ is 
\leqn{halliwell}{
H=\left[\Frac{V_0}{3-1/p}\right]^{\half}
\exp\{\Frac{\phi}{\sqrt{2p}}\},
}
that corresponds to the Halliwell attractor \cite{halliwell_paper}.
By using (\ref{conformal1}), (\ref{3metricconf}), (\ref{halliwell}) and 
with a choice of time parameter such that $\lim_{u\rightarrow 0}\phi=-\infty$,
the contribution of all the terms second order in
spatial gradients to the evolution of the conformal metric at this order is
\leqnarr{3conformalmetric}{
\dot{f}^{(3)}_{ij} & = & \Frac{-2c^{-2}}{(p+1)}\left[\Frac{V_0}{p(3p-1)}
\right]^{-p}u^{-2p+1}\tilde{R}^{(1)}_{ij} \nonumber \\
& + & \Frac{c^{-4}}{(1-p)}\left[
\Frac{V_0}{p(3p-1)}\right]^{-2p}u^{-4p+1}\left(-\half{\cal F}_{mn}
{\cal F}^{mn}f^{(1)}_{ij}+2{{\cal F}_i}^n{\cal F}_{jn}\right),\;p\not=1 
\label{3conformalmetric}
}
where $c$ is the integration constant and the seed metric is absorbed to
raise indices.
Integration of the above equation yields
\leqnarr{3conformalmetricintegrated}{
f^{(3)}_{ij} & = & \Frac{c^{-4}}{(1-p)(1-2p)}\left[\Frac{V_0}{p(3p-1)}
\right]^{-2p}u^{-4p+2}\left(-\quart{\cal F}_{mn}{\cal F}^{mn}f^{(1)}_{ij} 
+{{\cal F}_i}^n{\cal F}_{jn}\right)\nonumber \\
& + & \Frac{c^{-2}}{p^2-1}\left[\Frac{V_0}{p(3p-1)}\right]^{-p}
u^{-2p+2}\tilde{R}_{ij}+f^{(1)}_{ij}.
}
After taking (\ref{3metric}) and (\ref{3conformalmetric}) into account, 
the three-metric is given by
\leqnarr{3metricphi}{
\gamma^{(3)}_{ij}& = & \Frac{c^{-2}}
{(1-p)(1-2p)}\left[\Frac{V_0}{p(3p-1)}\right]^{-p}u^{-2p+2}
\left(-\quart{\cal F}_{mn}{\cal F}^{mn}f^{(1)}_{ij}+{{\cal F}_i}^n{\cal F}_{jn}
\right) \nonumber \\ 
& + & \Frac{1}{p^2-1}u^2\tilde{R}^{(1)}_{ij}+c^2\left[\Frac{V_0}{p(3p-1)}\right]
^pu^{2p}f^{(1)}_{ij}.
}
Two points regarding the above equation are in order.  First, as a perturbation
of a flat Friedmann-Robertson-Walker (FRW) model, the growth of 
inhomogeneities from terms second order in
spatial gradients
is encoded in $\gamma^{(3)}
_{ij}$.  One notices that with the choice of $u$ as the time parameter, such 
a growth resulting from the curvature is insensitive to the exact form of
the exponential potential.  In other words, deviations from a flat FRW cosmology
resulting from a magnetic field evolve quite differently from that of the 
inhomogeneities due to the space-time geometry.  In particular, one notices that if 
$p>1$, the contribution of a primordial magnetic field to spatial inhomogeneities decays rapidly.  

Secondly, the reader is cautioned
against a mini-superspace perturbation of a flat FRW model, namely, a 
perturbation within a mini-superspace cosmology.  For example, for a vanishing
magnetic field in a 
Bianchi {\rm I} cosmology where the conformal three-curvature is vanishing 
({\em i.e.}  $\tilde{R}_{ij}=0$), one would get the 
obviously incorrect result $\gamma^{(n)}_{ij}(t)=\gamma^{(1)}_{ij}(t)$
for $n\not=1$ (because $S^{(n)}=0$ for $n\not=0$).  The exact 
evolution equation for such a model could be derived from 
(\ref{momenta1}-\ref{momenta3}) and (\ref{fieldphi}-\ref{fieldA}) which 
could be solved numerically for generic initial conditions.  This problem
also exists in the dust model treated in PSS which has been quite successful
in deriving the contribution of short wavelength fields to structure formation
and the anisotropies of microwave background radiation.
\section{Conclusion}
The spatial gradient expansion of the generating functional was developed by
PSS to solve the Hamiltonian constraint in EHJ formulation of general 
relativity
for gravitationally interacting dust and scalar fields.  The spatial gradient 
expansion could be consistently applied to solve the Hamiltonian constraint
for gravitationally interacting electromagnetic and scalar fields.  At each 
order, the EHJ equation is a linear functional partial differential equation 
in 
the unknown functional $S^{(n)}$ which after a conformal transformation could be
integrated to yield $S^{(n)}$.  $S^{(2)}$ and $S^{(3)}$ were
calculated in detail and $S^{(4)}$ was given.  
Such an order-by-order solution of the EHJ equation
gives rise to order-by-order corrections to the fields evolving in a flat 
FRW model.  The corrections are due the presence of spatial inhomogeneities and
magnetic field.  Not surprisingly, such corrections start with terms second
order in spatial gradients.  The formalism was applied to the specific example
of a scalar field with potential $V=V_0{\rm exp}\{-\sqrt{\Frac{2}{p}}\phi\}$. 
Contributions
of all the terms second order in spatial gradients to the metric were 
derived.

I would like to thank Ali Mostafazadeh, Don Page and Dave Salopek for 
helpful discussions.

\end{document}